\documentclass[11pt]{article}

\usepackage{float}

\usepackage[preprint]{acl}

\usepackage{times}
\usepackage{latexsym}
\usepackage{amsfonts}
\usepackage{amsthm}
\usepackage{subcaption}
\usepackage[T1]{fontenc}
\usepackage[utf8]{inputenc}
\usepackage{amsmath,amsfonts,amssymb,amsthm,bm}

\usepackage{microtype}
\usepackage{inconsolata}
\usepackage{graphicx}
\usepackage{hyperref}
\usepackage{braket}

\title{Quantum Circuit Design using Complex valued Neural Network in Stiefel Manifold}

\author{
  Sayan Manna \\
  Department Artificial Intelligence \\ IIT Kharagpur \\
  \texttt{sayan2003@kgpian.iitkgp.ac.in} 
  \And
  Mahesh Mohan M R \\
  Department of Artificial Intelligence \\ IIT Kharagpur \\
  \texttt{mahesh.mohan@cai.iitkgp.ac.in} \\
}

\begin{document}
\maketitle
\begin{abstract}
Quantum algorithms operate on quantum states through unitary transformations in high-dimensional complex Hilbert space. In this work, we propose a machine learning approach to create the quantum circuit using a single-layer complex-valued neural network. The input quantum state is provided to the network, which is trained to approximate the output state of a given quantum algorithm. To ensure that the fundamental property of unitarity is preserved throughout the training process, we employ optimization in Stiefel Manifold. 
\end{abstract}

\section{Introduction}
Quantum computing exploits quantum mechanics to process information beyond the capabilities of classical computation \cite{nielsen2002}. A central research direction is the design of quantum algorithms with provable advantages over classical ones \cite{deutsch1992rapid,shor1994algorithms}. These are typically expressed in the \emph{quantum circuit model}, which are built from sequences of quantum gates acting on quantum states which are complex in nature \cite{barenco1995elementary,shende2006synthesis}, enabling practical implementation on quantum devices or simulators.

Despite progress, mapping quantum algorithms to suitable gate sequences remains challenging, with many approaches  \cite{ge2024quantum_synthesis_survey, shim2019connectivity_mapping_challenges} still lacking efficient solutions. This motivates automated methods for generating task-specific quantum circuits. At the same time, machine learning—well-established for prediction and classification—has increasingly been combined with quantum computing in recent research \cite{houssein2022machine}. 

In this paper, we propose a neural network-based approach for constructing quantum circuits. Existing methods such as Lie algebra approaches and Gram–Schmidt orthogonalization face limitations, as they cannot simultaneously guarantee unitarity preservation while ensuring a consistent decrease in the loss function. Here we have tried to solve this problem. Quantum evolution applies a unitary operator $U$ to a quantum state, where $U$ satisfies $U U^{\dagger} = I$, where $U^{\dagger}$ is conjugate transpose of $U$. Since quantum evolution is governed by unitary operators \cite{parker2025unitary}, we ensure that the network’s weight matrices remain unitary during training, enabling a natural mapping to quantum circuits. Our method leverages optimization of weight matrix of complex valued neural network in \emph{Stiefel manifold}, addressing limitations of existing approaches.

\section{Problem Definition}

Consider a quantum algorithm for which many input and output quantum states are known. Our objective is to design a neural network that can learn the corresponding unitary transformation of the quantum circuit. We construct a single-layer, complex-valued neural network for this purpose. For a circuit on $n$ qubits, the state vector lies in $\mathbb{C}^{2^n}$ and evolves as $\ket{\Psi'} = U \ket{\Psi}$ with $U \in \mathbb{C}^{2^n \times 2^n}$. Thus, the neural network takes $2^n$ complex inputs and is trained to learn the unitary transformation between input and output states. Let $W \in \mathbb{C}^{N \times N}$, $N = 2^n$ be the network weight matrix, and $f(W)$ a real-valued loss. Training becomes the constrained optimization
\begin{equation}
\min_{W \in \mathbb{C}^{N \times N}} f(W) 
\quad \text{subject to} \quad W^{\dagger} W = I
\end{equation}
i.e., optimizing $f(W)$ while preserving the unitarity of $W$.

\section{Literature Study}
Enforcing unitarity in neural network weights  has been studied in varius aspects \cite{arjovsky2016unitary,mhammedi2016orthogonal}. Use of machine learning to design scalable quantum circuit also has been studied in \cite{sarkar2024quantum,wan2016quantum}. Common approaches include Gram–Schmidt orthogonalization and Lie algebra parametrizations \cite{zomorodi2024optimal,coulaud2022orthogonalization}, where weights are re-orthogonalized after fixed epochs. However, these methods only enforce unitarity intermittently, often causing abrupt increases in loss and yielding matrices that are not unitary, limiting their physical realizability.
To address this, we adopt the \emph{Cayley transform} on the Stiefel manifold, first proposed for RNN training \cite{wisdom2016full}, which preserves unitarity at every update. This ensures smooth loss decrease and strictly unitary weights throughout training. We extend this idea to quantum circuit design, representing one of the first applications of Cayley-transform-based manifold optimization in quantum computing.

\section{Methodology}
The core of this approach relies on optimizing neural network weights over the Stiefel manifold. The Stiefel manifold, denoted as 
\(\mathcal{V}_{N}(\mathbb{C}^{N})\), is defined as the set of all complex \(N \times N\) matrices whose columns form an orthonormal basis in $\mathbb{C}^{N}$ \cite{tagare2011stiefel}, i.e.,
\begin{equation}
\mathcal{V}_{N}(\mathbb{C}^{N}) = \{ W \in \mathbb{C}^{N \times N} \; : \; W^{\dagger} W = I_{N} \}.
\label{eq:stiefel}
\end{equation}
Here, \(W^{\dagger}\) represents the conjugate transpose of \(W\), and \(I_{N}\) is the \(N \times N\) identity matrix. For any \(W \in \mathcal{V}_{N}(\mathbb{C}^{N})\), we consider a loss function \(f(W)\). The Euclidean gradient of \(f\) with respect to the entries of \(W\) is denoted by \(G \in \mathbb{C}^{N \times N}\), where 
$G_{ij} = \frac{\partial f}{\partial W_{ij}}$. From this gradient, we construct a skew-Hermitian matrix $A$ $(A^{\dagger}=-A)$ where,
\begin{equation}
A = G W^{\dagger} - W G^{\dagger}
\label{eq:skew}
\end{equation}
It defines a valid search direction tangent in the Stiefel manifold \cite{wisdom2016full}.

A descent curve along the manifold at training iteration \(k\) is then obtained by applying the Cayley transformation of \(A^{(k)}\) to the current solution \(W^{(k)}\). The update rule is expressed as:
\begin{equation}
W^{(k+1)} = \left( I + \frac{\lambda}{2} A^{(k)} \right)^{-1} 
\left( I - \frac{\lambda}{2} A^{(k)} \right) W^{(k)}
\label{eq:update}
\end{equation}
where \(\lambda \in \mathbb{R}\) is the learning rate, chosen such that \(\lambda > 0\) and sufficiently small. This formulation guarantees two important desired properties:  
\begin{enumerate}
    \item The updated weight matrix remains on the Stiefel manifold, i.e., \(W^{(k+1)} \in \mathcal{V}_{N}(\mathbb{C}^{N})\).  
    \item The loss function decreases globally, i.e., \(f(W^{(k+1)}) \leq f(W^{(k)})\).  
\end{enumerate}
Please see the detailed proof of these properties in the Appendix.

We design a single-layer complex-valued neural network with weights initialized as a random unitary matrix (\(W \in \mathcal{V}_{N}(\mathbb{C}^{N})\)). Training is performed using the prescribed update rule, yielding a final unitary matrix $W_{\text{final}}\in \mathcal{V}_{N}(\mathbb{C}^{N})$. After the network training, we use the transpile function to convert the $W_{\text{final}}$ into quantum circuits. 

The \texttt{transpile} function converts a quantum circuit into an optimized form executable on specific hardware or simulators (e.g, Qiskit’s \texttt{transpile} function).

\section{Results and Discussion}
We used a 5 qubit standard quantum circuit for experiment.
\vspace{-0.5cm}
\begin{figure}[H] 
    \centering
    \begin{subfigure}[b]{0.48\columnwidth}
        \centering
        \includegraphics[width=\linewidth]{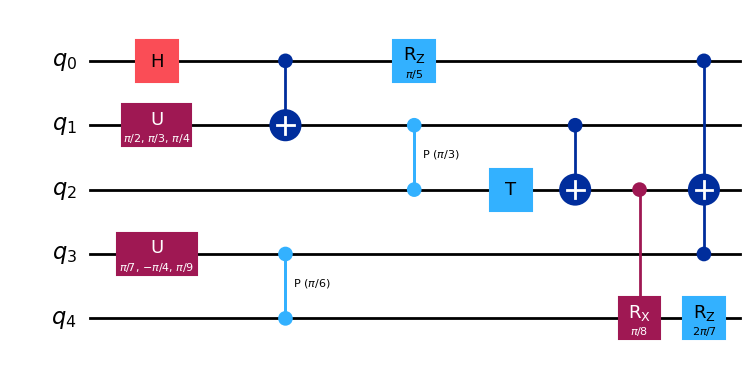}
        \caption{5-qubit  sample quantum circuit}
        \label{fig:loss1}
    \end{subfigure}
    \hfill
    \begin{subfigure}[b]{0.48\columnwidth}
    \vspace{0.5cm}
        \centering
        \includegraphics[width=\linewidth]{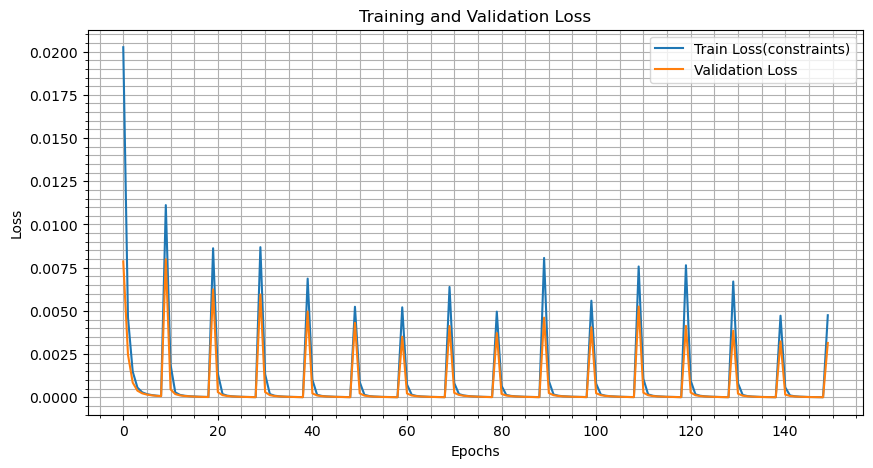}
        \caption{Loss vs Epochs using Gram-Schmidt method}
        \label{fig:gram}
    \end{subfigure}
    \caption{Sample circuit and Gram-Schmidt loss}
    \label{fig:training}
\end{figure}
As shown in Fig.~\ref{fig:gram}, the training loss increases whenever the Gram-Schmidt orthogonalization constraint is being applied to the weight matrix. Here unitarity is not being maintained during training.
\begin{figure}[H] 
    \centering
    \begin{subfigure}[b]{0.48\columnwidth}
        \centering
        \includegraphics[width=\linewidth]{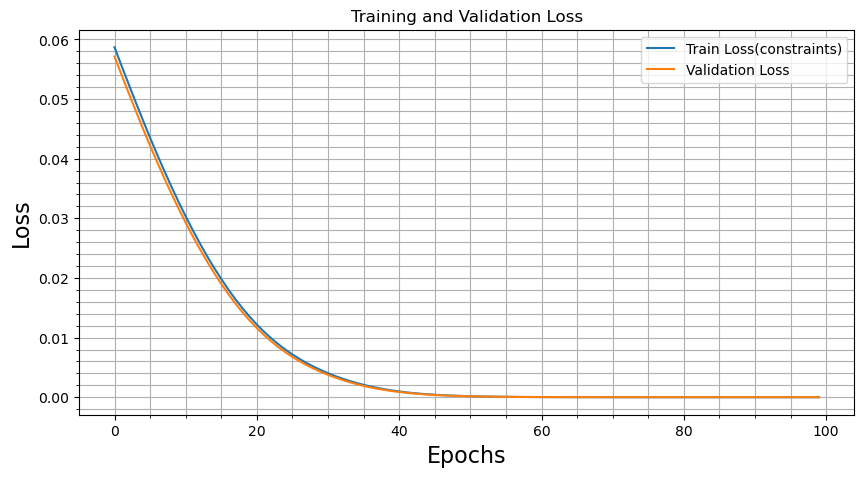}
        \caption{Loss vs Epochs using optimization on Steifel manifold}
        \label{fig:loss0}
    \end{subfigure}
    \hfill
    \begin{subfigure}[b]{0.48\columnwidth}
        \centering
        \includegraphics[width=\linewidth]{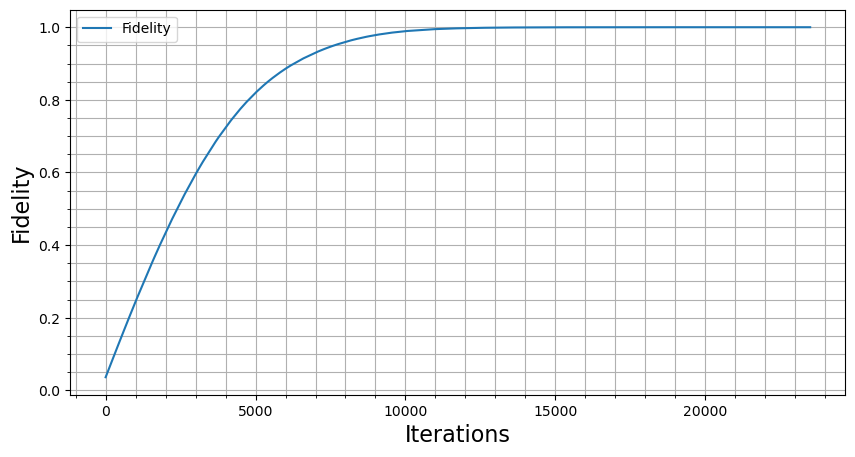}
        \caption{Fidelity vs iteration on Stiefel manifold}
        \label{fig:fidelity0}
    \end{subfigure}
            \caption{Training loss and Fidelity}

\end{figure}
According to our approach as shown in Fig.~\ref{fig:loss0} the loss is decreasing when Eq.(\ref{eq:update}) is being used. Throughout the training, the unitary error is observed to be of the order 
$\left\| W W^{\dagger} - I \right\|_2^2 \sim 10^{-11}$. This shows that unitarity is being maintained. We quantify the closeness between the learned unitary $U_{\text{learned}}$ and the target unitary 
$U_{\text{true}}$ using the fidelity
\begin{equation}
F = \frac{1}{d} \left|\mathrm{Tr}\!\left(U_{\text{true}}^\dagger U_{\text{learned}}\right)\right|
\end{equation}
where $d$ is the dimension of the uniatry matrix. A value $F=1$ indicates $U_{\text{learned}}=U_{\text{true}}$. As shown in Fig.~\ref{fig:fidelity0}, the fidelity increases during training and finally reaches $F=1$.

\nocite{*}
\bibliographystyle{plain}
\bibliography{custom}
\appendix
\section*{Appendix}
\subsection*{Optimization in Stiefel Manifold}
\renewcommand{\theequation}{A\arabic{equation}}
\setcounter{equation}{0} 
We consider the Stiefel manifold of all $N \times N$ complex-valued matrices whose columns are $N$ orthonormal vectors in $\mathbb{C}^N$. It is defined as
\[
\mathcal{V}_N(\mathbb{C}^N) = \{ W \in \mathbb{C}^{N \times N} \;|\; W^\dagger W = I \}.
\]

\subsubsection*{Tangent Space}
A tangent space at a point on a manifold is the set of all possible directions in which we can move without leaving the manifold.  
Consider $W \in \mathcal{V}_N(\mathbb{C}^N)$. Since every point on the Stiefel manifold represents a unitary matrix, the tangent space at $W$, written as $\mathcal{T}_W \mathcal{V}_N(\mathbb{C}^N)$, contains all the directions (matrices) in which we can move infinitesimally while staying on the manifold.

The tangent space at $W$ is defined as
\begin{equation}
  \mathcal{T}_W \nu_N(\mathbb{C}^N) = \{ Z \in \mathbb{C}^{N \times N} \;|\; W^\dagger Z + Z^\dagger W = 0 \}.  
\end{equation}

This condition ensures that if we move a small step $\epsilon$ to $W + \epsilon Z$, we still satisfy unitarity up to first order.

\subsubsection*{Proof}
\vspace{-0.5cm}
\begin{align*}
& (W+ \epsilon Z)(W+ \epsilon Z)^\dagger \\
&= (W+ \epsilon Z)(W^\dagger + \epsilon Z^\dagger) \\
&= W W^\dagger + \epsilon Z W^\dagger + \epsilon W Z^\dagger + O(\epsilon^2) \\
&= I + \epsilon \big( ZW^\dagger + WZ^\dagger \big) + O(\epsilon^2).
\end{align*}
Thus, if $ZW^\dagger + WZ^\dagger = 0$, we have $(W+ \epsilon Z)(W+ \epsilon Z)^\dagger = I$ up to first order, showing $Z \in \mathcal{T}_W \mathcal{V}_N(\mathbb{C}^N)$.

\subsubsection*{Skew-Hermitian Representation}
Now, let $A$ be any skew-Hermitian matrix, i.e., $A^\dagger = -A$. Then
\begin{equation}
    Z = AW \in \mathcal{T}_W \mathcal{V}_N(\mathbb{C}^N)
\end{equation}

is a valid tangent vector. Indeed,
\begin{align*}
& (W+ \epsilon Z)(W+ \epsilon Z)^\dagger \\
&= (W+ \epsilon AW)(W+ \epsilon AW)^\dagger \\
&= (W+ \epsilon AW)(W^\dagger - \epsilon W^\dagger A) \\
&= WW^\dagger + \epsilon AW W^\dagger - \epsilon W W^\dagger A + O(\epsilon^2) \\
&= I + \epsilon A - \epsilon A + O(\epsilon^2) \\
&= I + O(\epsilon^2).
\end{align*}
Hence, $Z = AW$ indeed lies in the tangent space at $W$.

\subsubsection*{Canonical Inner Product and Riemannian Gradient}
Next, we define the \textit{canonical inner product}.  
The Stiefel manifold becomes a Riemannian manifold by introducing an inner product in its tangent spaces.  
Let $Z_1, Z_2 \in \mathcal{T}_W \mathcal{V}_N(\mathbb{C}^N)$, then the canonical inner product is given by
\begin{equation}
\langle Z_1, Z_2 \rangle_c = \mathrm{tr} \left( Z_1^{\dagger} \left(I - \tfrac{1}{2} W W^{\mathrm{H}} \right) Z_2 \right).
\end{equation}

Next, we define the gradients.  
Consider the smooth loss function
\[
f(W) : \mathbb{C}^{N \times N} \to \mathbb{R}.
\]
The Euclidean gradient $G \in \mathbb{C}^{N \times N}$ is defined as
\begin{equation}
G_{ij} = \frac{\partial f}{\partial W_{ij}}.
\end{equation}

If $Z \in \mathbb{C}^{N \times N}$, the differential of $f$, denoted by $Df(W)[Z] : \mathbb{C}^{N \times N} \to \mathbb{C}$,  
gives the derivative of $f$ in the direction $Z$ at $W$:
\begin{equation}
Df(W)[Z] = \mathrm{tr}(G^{\dagger} Z) 
= \sum_{i,j} \frac{\partial f}{\partial W_{ij}} \, Z_{ij}.
\end{equation}

Now we define the Riemannian gradient of $f$, denoted $\nabla_c f \in \mathcal{T}_W \mathcal{V}_N(\mathbb{C}^N)$.  
It is obtained as the projection of $G$ onto the tangent space:
\begin{equation}
\nabla_c f = A W, \quad \text{where } A = G W^{\dagger} - W G^{\dagger}.
\end{equation}

Under the canonical inner product, the vector $AW$ with  $A = G W^{\dagger} - W G^{\dagger}$ represents the action of $Df(W)$ on the tangent space $\mathcal{T}_W \mathcal{V}_N(\mathbb{C}^N)$.

Consider the curve
\begin{equation}
    Y(\lambda) = \left(I + \tfrac{\lambda}{2} A \right)^{-1} \left(I - \tfrac{\lambda}{2} A \right) W,
\end{equation}
where $W \in \mathcal{V}_N(\mathbb{C}^N)$, $G = \nabla f(W)$ is the Euclidean gradient of $f$, and $A = GW^\dagger - WG^\dagger, \qquad \lambda \in \mathbb{R}, \, \lambda>0 \text{ (sufficiently small)}$.\\

\textbf{Claim 1:} $Y(\lambda) Y(\lambda)^\dagger= I$, i.e., $Y(\lambda) \in \mathcal{V}_N(\mathbb{C}^N)$.

\textbf{Proof:}  
Since
\begin{align*}
    A^\dagger &= (GW^\dagger - WG^\dagger)^\dagger \\
              &= WG^\dagger - GW^\dagger \\
              &= -A,
\end{align*}
$A$ is a skew-Hermitian matrix.\\

Now,
\begin{align*}
    & Y(\lambda)^\dagger Y(\lambda)\\
    &= \Big( \big(I + \tfrac{\lambda}{2} A \big)^{-1} \big(I - \tfrac{\lambda}{2} A \big) W \Big)^\dagger 
       \big(I + \tfrac{\lambda}{2} A \big)^{-1}\\
      & \hspace{5cm} \big(I - \tfrac{\lambda}{2} A \big) W \\
    &= W^\dagger \big(I - \tfrac{\lambda}{2} A \big)^\dagger 
       \big( (I + \tfrac{\lambda}{2} A )^{-1} \big)^\dagger 
       \big(I + \tfrac{\lambda}{2} A \big)^{-1}\\
       & \hspace{5cm}\big(I - \tfrac{\lambda}{2} A \big) W\\
    &= W^\dagger \big(I + \tfrac{\lambda}{2} A \big) 
       \big(I - \tfrac{\lambda}{2} A \big)^{-1} 
       \big(I + \tfrac{\lambda}{2} A \big)^{-1}\\ 
       & \hspace{5cm}\big(I - \tfrac{\lambda}{2} A \big) W \\
    &= W^\dagger \big(I + \tfrac{\lambda}{2} A \big) 
       \big( (I + \tfrac{\lambda}{2} A)(I - \tfrac{\lambda}{2} A) \big)^{-1} 
       \big(I - \tfrac{\lambda}{2} A \big) W \\
    &= W^\dagger \big(I + \tfrac{\lambda}{2} A \big) 
       \big(I - \tfrac{\lambda^2}{4} A^2 \big)^{-1} 
       \big(I - \tfrac{\lambda}{2} A \big) W \\
    &= W^\dagger \big(I + \tfrac{\lambda}{2} A \big) 
       (I + \tfrac{\lambda}{2} A)^{-1} (I - \tfrac{\lambda}{2} A)^{-1} 
       \big(I - \tfrac{\lambda}{2} A \big) W \\
    &= W^\dagger I W \\
    &= W^\dagger W \\
    &= I.
\end{align*}

Therefore,
\begin{equation}
    Y(\lambda)^\dagger Y(\lambda) = I 
    \quad \Longrightarrow \quad 
    Y(\lambda) \in \mathcal{V}_N(\mathbb{C}^N).
\end{equation}

\textbf{Claim 2:}
For small enough $\lambda$, we have 
\[
f(Y(\lambda)) \leq f(W).
\]

\textbf{Proof:} 
We have
\[
Y(\lambda) = \left(I + \tfrac{\lambda}{2} A \right)^{-1} \left(I - \tfrac{\lambda}{2} A \right) W.
\]
At $\lambda = 0$, clearly $Y(0) = W$.  

We know that if $M(\lambda)$ is invertible and differentiable, then
\begin{equation}
  \frac{d}{d\lambda} M(\lambda)^{-1} 
= - M(\lambda)^{-1} \frac{d M(\lambda)}{d\lambda} M(\lambda)^{-1}. 
\label{eq:derivative}
\end{equation}

Now,
\begin{align*}
        & Y'(\lambda)\\
        & = \Bigg[ \frac{d}{d\lambda} \Big(I + \tfrac{\lambda}{2} A \Big)^{-1} 
\Big(I - \tfrac{\lambda}{2} A \Big) 
+\\
& \hspace{2cm} \Big(I + \tfrac{\lambda}{2} A \Big)^{-1} \frac{d}{d\lambda} \Big(I - \tfrac{\lambda}{2} A \Big) \Bigg] W
\tag{A10}
\end{align*}

Using equation~\eqref{eq:derivative}, we obtain
\[
Y'(0) = \Big(-\tfrac{1}{2} A \Big) I W + I \Big(-\tfrac{1}{2} A \Big) W
= -AW
\]

Thus, the tangent vector of the curve $Y(\lambda)$ at the starting point is
\[
Y'(0) = -AW.
\]

Now, under the canonical inner product on the tangent space, the gradient in the Stiefel manifold of the loss function $f$ with respect to the matrix $W$ is $AW$, where $A = GW^\dagger - WG^\dagger$. That is, for every tangent vector $Z$ (or in the direction of $Z$),
\[
Df(W)[Z] = \langle AW, Z \rangle_c \tag{A11}
\]

Therefore, the directional derivative of $f$ along the curve at $\lambda = 0$ is
\begin{align*}
     \frac{d}{d\lambda}\Big|_{\lambda = 0} f(Y(\lambda)) 
&= Df(W)[Y'(0)]\\
&= Df(W)[-AW]\\
&= \langle AW, -AW \rangle_c\\
&= - \langle AW, AW \rangle_c\\
&= - \| AW \|_c^2 \leq 0
\end{align*}
\begin{equation}
         \frac{d}{d\lambda}\Big|_{\lambda = 0} f(Y(\lambda)) = - \| AW \|_c^2 \leq 0
         \label{first_derivative} \tag{A12}
\end{equation}

Hence, the first directional derivative is strictly negative if the Riemannian gradient is non-zero (i.e., $AW \neq 0$).

Because $f$ is smooth, for small $\lambda> 0$ we have the Taylor expansion with remainder,
\begin{equation}
    f(Y(\lambda)) = f(W) + \lambda \frac{d}{d\lambda}\Big|_{\lambda = 0} f(Y(\lambda)) + R(\lambda)\tag{A13}
\end{equation}

where $R(\lambda) = O(\lambda^2)$.

More concretely, there exist $c > 0$ such that  
\begin{equation}
    |R(\lambda)| \leq c \lambda^2 \tag{A14}
\end{equation}

Therefore, for $\lambda > 0$, using ~\eqref{first_derivative}, we obtain
\begin{align*}
    & f(Y(\lambda)) \leq f(W) - \lambda \| AW \|_c^2 + c \lambda^2 \\
& = f(W) - \lambda \| AW \|_c^2 \left( 1 - \frac{c\lambda}{\| AW \|_c^2} \right)
\end{align*}

If $AW \neq 0$, pick $\lambda > 0$ small enough so that
\[
\frac{c \lambda}{\| AW \|_c^2} \leq \frac{1}{2}
\]

Then, for such $\lambda$,
\[
f(Y(\lambda)) \leq f(W) - \frac{\lambda}{2} \| AW \|_c^2 \leq f(W) \tag{A15}
\]

Thus, for all sufficiently small positive $\lambda$, the update rule decreases $f$ unless $AW = 0$.
\subsection*{Additional Result: 2-qubit quantum circuit synthesis}
In this example, we take 2-qubit entanglement operation. Now we want the quantum circuit for it using neural network. The input and output quantum states of the quantum operation is known. First, we create a complex valued neural network  with 4 inputs and 4 outputs which ensures the dimension of unitary matrix to be $4 \times4$. We train the network with Caley update rule and get the following results.
\begin{figure}[H] 
    \centering
    \begin{subfigure}[b]{0.48\columnwidth}
        \centering
        \includegraphics[width=\linewidth]{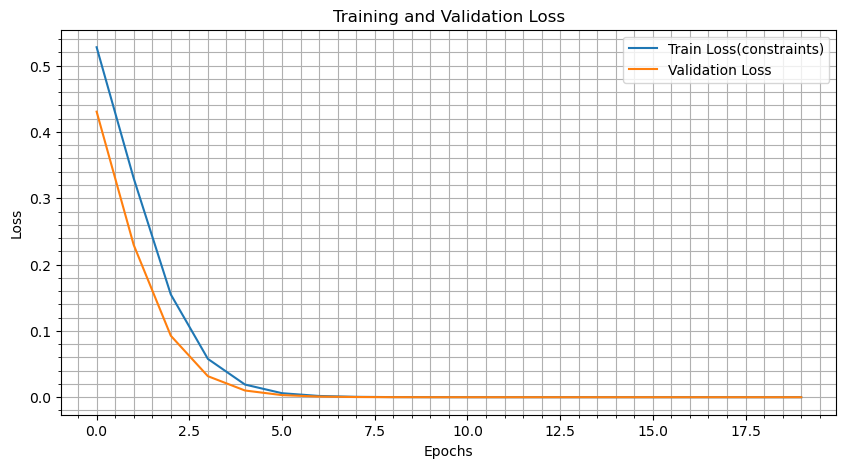}
        \caption{Loss vs Epochs in steifel manifold}
        \label{fig:loss}
    \end{subfigure}
    \hfill
    \begin{subfigure}[b]{0.48\columnwidth}
        \centering
        \includegraphics[width=\linewidth]{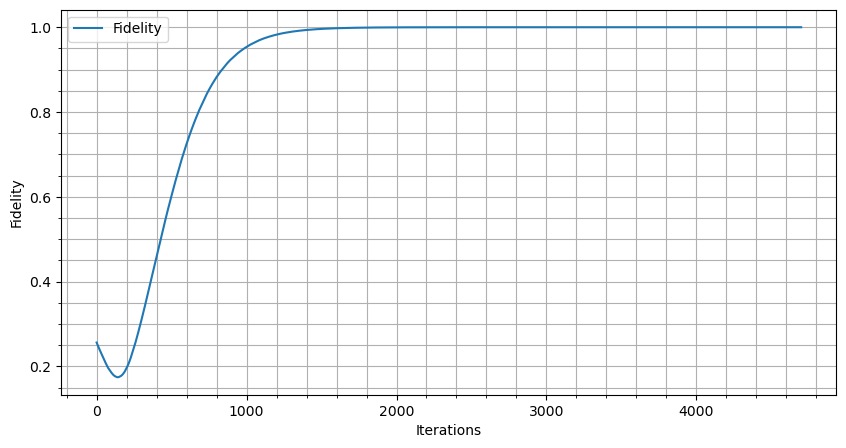}
        \caption{Fidelity vs iteration in Stiefel Manifold}
        \label{fig:fidelity}
    \end{subfigure}
            \caption{Training loss and Fidelity for a 2-qubit quantum operation}
\end{figure}
As we can see, the loss is decreasing and fidelity is reaching 1. Next, we extract the unitary weight matrix from the neural network and the obtained unitary error, $\left\| W W^{\dagger} - I \right\|_2^2 = 5.637545\times10^{-14}$ which is negligible. Then we transpile the unitary matrix with Qiskit's Transpile function to obtain the corresponding quantum circuit.
\begin{figure}[H]
    \centering
    \includegraphics[width=0.5\textwidth]{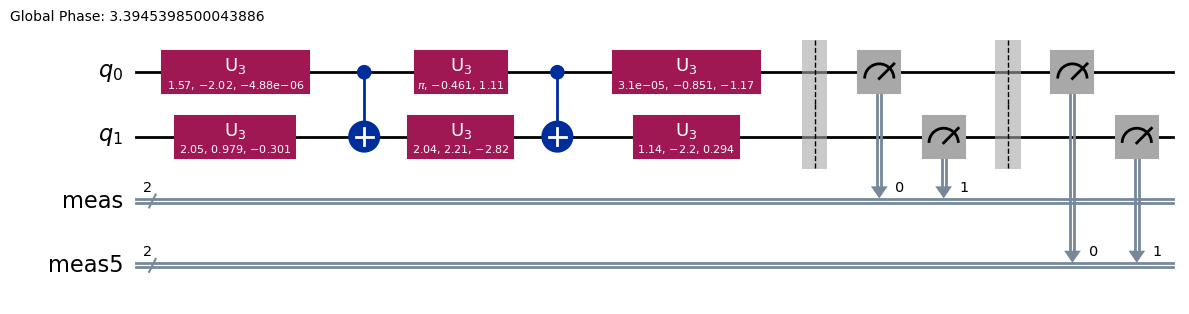} 
    \caption{2-qubit optimized quantum circuit after transpilation}
    \label{fig:myfigure2}
\end{figure}

\end{document}